\documentstyle[12pt]{article}
\begin{document}

\title{ Self-organization, ergodicity breaking, phase
transition and synchronization in two-dimensional traffic-flow model}
\author{Yu Shi\thanks{electronic address: yushi@fudan.ihep.ac.cn}\\
Technology (World Laboratory),\\ P. O. Box 8730,
Beijing 100080, People's Republic of China\\ and\\
Fudan-T. D. Lee Physics Laboratory and Department of Physics,
\\Fudan University, Shanghai 200433,
People's Republic of China\thanks{mailing address}}
\maketitle
\newpage
\begin{abstract}
Analytical investigation is made on the
two-dimensional traffic-flow model with alternative movement and
exclude-volume effect between right and up arrows [Phys. Rev. {\bf A}
46
R6124 (1992)]. Several exact results are obtained, including the
upper critical density above which there are only jamming
configurations,
and the lower critical density below which there are only moving
configurations. The observed jamming transition takes place at
another
critical density $p_{c}(N)$, which is in the intermidiate region
between the lower and upper critical densities. It is derived that
$p_{c}(N)\,=\,CN^{\alpha}$, where $C$ and $\alpha$ are determined to
be
respectively  $0.76$ and $-0.14$    from previous numerical
simulation.
This transition is suggested to be a second-order phase transition,
the order parameter is found.
The nature of
self-organization, ergodicity breaking and synchronization are
discussed,
Comparison with the sandpile model is made.
\end{abstract}
\vspace{1.5cm}
PACS numbers: 05.70.Ln, 64.60.Cn, 64.60.Ak, 89.40.+k

\newpage
\section*{I. INTRODUCTION}

        Much attention has been paid on cellular automaton models to
investigate
complex systems.  These models can be viewed as statistical models
with
dynamics. Recently some
models approaching traffic-flow problems have been studied.
There are mainly two classes of models.
The model of one class, introduced by Nagel and
Schrekenberg [1-2], is defined on a one-dimensional array or simple
network, where each site can be in one of several states representing
emptiness or occupied by a car  with one of the possible velocities.
The iterative rules include accelaration, deceleration, randomization
and
movement. Indications were found near the point of maximum throughout
for a phase transition  separating low-density lameller flow from
high-density
jammed behavior. Completely different from this model,  the model of
another
class, introduced by Biham, Middleton and Levin (BML) \cite{biham},
is defined
on a square lattice with periodic boundary condition. Each site
contains
either an arrow pointing upwards, an arrow pointing to the right, or
is empty.
The dynamics is controlled by the traffic light, such that the right
arrows
move only on even time steps and the up arrows move on odd time
steps. On
even
time steps, each right arrows moves one lattice constant to the right
unless
the site on its righhand side is occupied, either up or right. If
it is blocked by another arrow it does not move, even if during the
same time step the blocking arrow moves out of that site. Similar
rules
apply to the up arrows, which move upwards. The velocity $v$ of {\em
an}
arrow is defined to be the number of success moves it makes within  a
time interval of its turn $\tau$, which is just the number of odd or
even
time steps within a ``real'' time interval, be divided by
$\tau$ \cite{exp}. It has maximal value $v\,=\,1$, indicating that
the arrow
was never blocked, while $v\,=\,0$ means that  the arrow was stopped
for the
entire duration. The average velocity $\overline{v}$
for the system is obtained by
averaging $v$ over {\em all} the arrows in the system.

        The BML model is fully deterministic. It is called  to be
self-organized because whatever the initial configuration of the
system
is, in the asymptotic configurations, {\em all} the arrows move
freely in
their turn
hence the velocity averaged over all the arrows is
$\overline{v}\,=\,1$,
or they
are {\em all} stucked. These two kinds of configurations are referred
to as moving and jamming ones, respectively. In the language of
dynamics,
there are  two and only two  kinds of attractors.
Arriving finally which of the two  depends on
both the density of arrows and the initial configuration.
This essence escaped the attention in previous analytical attempts
\cite{nagatani}.
In order to show that the critical density above which
all arrows are stuck is less
than $1$, Chau, Hui and Woo \cite{chau}
find certain jamming configurations with density
less than $1$.
But it was not proved that
{\em all} the configurations of the same density
are jamming. In fact, as has been
indicated in Ref.  \cite{biham}, the
density for a configuration to be jamming can be as small as $2/N$,
where
$N$ is lattice size, but this case is atypical.
In a more complicated
model \cite{freund}, more realistic rules and results are adopted. We
think that
the most important factor leading to jamming transition are also the
exclude-volume effect between cars (arrows) of different type, as in
the
BML model.

We may ask,
is there any
configuration in which some arrows are moving while others are
blocked? The answer is yes. Consider a column whose two neighboring
columns
are all full of up arrows,
this column is occupied by up arrows less than $N/2$.
Then asymptotically the arrows in this column are moving forever
independent
of other arrows, which may be all blocked. But such configurations
are
very few, {\it i.e.}, its  volume in the phase space is negligible
compared
with that of
moving or
jamming configurations, especially when we consider asymptotic
configurations
after self-organization. Therefore one can only consider
moving and jamming configurations.

Here we give the complete picture for asymptotic configurations,
neglecting those with both moving and blocked arrows.
For very low density, there are only moving configurations. For very
high
density, there are only jamming configurations. Therefore there are a
lower
critical density and an upper critical density. Between these two
critical
densities, the asymptotic configuration can be moving or jamming,
dependent
on the initial configurations. As will be discussed, there is another
critical density above which the asymptotic configurations are
typically
jamming with moving one negligible. In fact this is just the jamming
transition uncovered in simulation  in Ref. \cite{biham}.
The contents of this article is as follows. For convenience of
discussions we
introduce some notations in Sec. II.
In Secs. III and IV, some exact results are given. The necessay
conditions for the
system to be moving or jamming are discussed,
thus the upper and lower critical
densities are determined exactly. In Sec. V, by
considering the typical
pattern formation of the jamming cluster, we obtain the
third critical density above which the observed jamming transition
takes place,
the dependence on
the lattice size is determined. This transition is suggested by sound
reason to be  a
second-order phase transition, the order parameter is found.
Some discussions concerning
the nature of self-organization, ergodicity
breaking and synchronization are
made in Sec. VI, where we also compare the BML model with the
sandpile
model.

\section*{II. NOTATIONS}

For convenience of discussions, some notations are used throughout
this
article. The lattice is $N\times N$, the density of up (right)
arrows is $p_{\uparrow}\,=\,n_{\uparrow}/N^{2}$
($p_{\rightarrow}\,=\,n_{\rightarrow}/N^{2}$), where $n_{\uparrow}$
($n_{\rightarrow}$) is the number of up (right) arrows. The total
density
of arrows
is $p\,=\,p_{\uparrow}\,+\,p_{\rightarrow}$. The number of empty
lattice
points is denoted as $n_{0}$.
The empty sites can be treated as occupied by holes, which are able
to move.
Each lattice point is given
a coordinate $(i,j)$, where $i$ and $j$ are $x$ and $y$ coordinates,
respectively. The lower-left corner is $(1,1)$. The periodic boundary
condition can be expressed as
\begin{equation}
(i+N,j)\,=\,(i,j+N)\,=\,(i+N,j+N)\,=\,(i,j).
\end{equation}
This periodicity makes every lattice point equivalent
(when there is no arrow), similar to (but
different from) that on a sphere. As will be used in later Sections,
the appearance of the lattice can be transformed without any real
change,
for example,  so that it can be seen that
the lattice can also be
viewed as a  parallelogram  also with the periodic boundary condition.
This parallelogram   is made up of  $N$ lines parallel to the left-
falling
diagonal
of the original square, on each of these lines
there are also
$N$ lattice pionts. Hence we say that the lattice is composed of
$N$ left-falling diagonals.  The lattice points on the line linking
$(1,i)$ and $(i,1)$ and on the line linking $(i+1,N)$ and $(N,i+1)$
belong to a same left-falling diagonal. Similarly, the lattice can
also
be viewed as being
made up  of right-falling diogonals. Since the arrows
are right or up, the former
viewpoint will be used in studying the moving configurations.

To avoid confusion, the word ``state'' is used for the lattice
points, while
``configuration'' is for the whole system. The state of $(i,j)$ is
denoted as $|i,j>$. $|i,j>\,=\,\uparrow,\,\rightarrow$ or $0$ if
 $(i,j)$ is occupied by an up arrow, a right arrow or is empty,
respectively.
 $|i,j>$ is, of course, dependent on time, so it can be written as
 $|i,j>(t)$ when needed, here $t$ is the corresponding  one for the
 given arrow. It is obvious that
 $|i,j>(t)\,=\,|i+\delta,j>(t+\delta)$ if $|i,j>(t)\,=\,\rightarrow$,
 $|i,j>(t)\,=\,|i,j+\delta>(t+\delta)$ if $|i,j>(t)\,=\,\uparrow$.
The case for $|i,j>\,=\,0$ will be discussed in the next Section.

\section*{III. EXACT RESULTS ON MOVING CONFIGURATION}
First we point out that not only jamming configuration, but also
moving
configuration is stationary, since all the same kind of
arrows move simultaneously
thus form a rigid body.
Considering the sequential arrangement for different kinds of
arrows, we may think
that the whole system moves  freely as  a rigid body. In fact
if we make a Galilean transformation to let one kind of arrow static,
the
system of the other kind of arrow hops between two position all the
time.
  In the phase space, the system hops between two fixed points.
  The exact results are stated in the form of lemma and theorem.

{\em Lemma 1.-In a moving configuration, at any time, if there
is an up (right) arrow on  a left-falling diagonal,
then any other lattice points on this   left-falling diagonal are
   also occupied by an up (right) arrow or is empty.}

{\em Proof.}-Assume $|i,j>\,=\,\uparrow$, while
$|i-\delta, j+\delta>\,=\,\rightarrow$, where $\delta$ is a positive
integer.
If it is on an odd (even) time step, then
after $\delta$ time steps in the turns for
up (right) arrow,
the right (up) arrow will be forbidden to move by the up (right)
arrow.
Because of periodic boundary condition,
every lattice point other than $(i,j)$ on the same
left-falling diagonal is $(i-\delta, j+\delta)$ with $\delta\,>\,0$.
Therefore there
can not be both up arrows and right arrows on a same left-falling
diagonal.  Q.E.D

    From this lemma, we know that  all  of the
left-falling diagonals with the same kind of arrows and holes,
form a rigid mody moving freely.
     So for $|i,j>(t)\,=\,0$,
     $|i,j>(t)\,=\,|i,j+\delta>(t+\delta)$ if it
     belongs to a left-falling diagonal with up arrows,
     while
     $|i,j>(t)\,=\,|i+\delta,j>(t+\delta)$ if it
     belongs to a left-falling diagonal with right arrows.

{\em Lemma 2.-In a moving configuration, at any time, there is at
least one
left-falling diagonal without any arrows, i.e., only made up of
holes.}

{\em Proof.}-Without lose of generality, consider  an odd time step.
For a left-falling
diagonal
of up arrows, its upside left-falling diagonal cannot be that of
right arrows
even if these right arrows are not in the same columns of those up
arrows,
or there will be a
left-falling diagonal where there are both up and right arrows, which
is forbidden by lemma $1$.

There must be at least one left-falling diagonal of up arrows whose
upside left-falling diagonal is
not that of up arrows, or there are only up arrows on the  whole
lattice.
 This upside left-falling diagonal should be empty.

 On an odd time step, for a left-falling diagonal of right arrows,
 it is unnecessary for its righthand side
 left-falling diagonal, which is just the upside one, to be empty. If
 the righthand side left-falling diagonal
 is that of up arrows, it will leave an empty left-falling diagonal
 after its movement.
 Therefore the least number of empty left-falling diagonal is $1$.
Q.E.D.

{\em Theorem 1.-For $N\,>\,2$, the necessary condition of  formation
of
any possible moving configuration is $p\,\leq\,1/2+(N-4)/2N^{2}$ if
$N$ is
even, $p\,\leq\,1/2-1/N^{2}$ if $N$ is odd. This is also
sufficient for that  moving configuration can form.}

{\em Proof.}-It is obvious that there cannot be any moving
configuration
with both kinds of arrows for
$N\,=\,2$, hence $N\,>\,2$ is assumed in the following.
In the connected left-falling diagonals with the same kind of arrows,
any up (right) arrow on
the upside (righthand side) left-falling diagonal should be on
the upside (righthand side) of
an empty site, or it is not moving configuration.
For even number of connected left-falling diagonals with the
same kind of arrows,
the number of arrows is at most equal to the number of empty sites.
For
odd number of connected left-falling diagonals with the same
kind of arrows, there can be at most
$(N-1)-1$ more than empty sites in addition. To make the
number of empty left-falling diagonals be the least value $1$,
all the left-falling diagonals with the
same kind of arrows must be connected. So if $N$ is even, the number
of the
left-falling diagonals with up arrows and that of left-falling
diagonals with right arrows can be
all odd, hence
\begin{equation}
n_{\uparrow}\,\leq\,n_{0}^{(1)}+(N-2),
\label{up}
\end{equation}
while
\begin{equation}
n_{\rightarrow}\,\leq\,n_{0}^{(2)}+(N-2),
\label{right}
\end{equation}
where $n_{0}^{(1)}$ ($n_{0}^{(2)}$) is the number of empty sites in
left-falling diagonals
with up (right) arrows.
If $N$ is odd, one of the numbers of the left-falling diagonals with
the same kind of
arrows is even while another is odd. Hence either Eq.\ (\ref{up}) is
valid
while
\begin{equation}
n_{\rightarrow}\,\leq\,n_{0}^{(2)},
\end{equation}
or
Eq.\ (\ref{right}) is valid while
\begin{equation}
n_{\uparrow}\,\leq\,n_{0}^{(1)}.
\end{equation}
Combining with
\begin{equation}
n_{0}\,\geq\,n_{0}^{(1)}\,+\,n_{0}^{(2)}\,+N,
\end{equation}
we obtain
\begin{equation}
p\,\leq\,\frac{1}{2}+\frac{N-4}{2N^{2}}
\end{equation}
if $N$ is even, and
\begin{equation}
p\,\leq\,\frac{1}{2}-\frac{1}{N^{2}}
\end{equation}
if $N$ is odd.
Q.E.D.

This theorem shows that the argument that the largest
possible $p$ for moving configuration is $2/3$ \cite{biham} is false.

\section*{IV. EXACT RESULTS ON JAMMING CONFIGURATION}
Since an up arrow can only be blocked by the upside arrow, which can
be
up or right one, while a right arrow can only be blocked by the right
side
arrow, these arrows thus form a directed path in a jamming
configuration.
The directions of
all the directed path here are rightward or upward. Considering the
periodic boundary coundition, it is easy to know
the following lemma 3.

{\em Lemma 3.-In the jamming configuration, starting from an
arbitrary arrow,
one can obtain  a directed
path which return to either this starting arrow or
an arrow in this path.}

Such a path can be called a closed path. If it return to the starting
arrow,
it is referred to as a circular path. of course, whenever there is a
closed path, there is a circular path, which is a part of the former.

{\em Lemma 4.-In the jamming configuration, there must be at least
one circular path.}

This is just the necessary condition of for a configuration to be
jamming.

{\em Lemma 5.-The length of a circular path is  $N$ if it is made up
of
only one kind of arrows, while is $2N$ if made up of both kinds. Here
the length is defined as the number of lattice points.}

{\em Proof.}-The first half statement is obvious
since the  circular path made up of one kind of arrows is
parallel to the edge of the square. If the circular path is made up of
both kinds of arrows, because  it is directed, generally
it  appears as two part in the square lattice, for
example,  one part is a directed path connecting
$(1,J)$ and $(I,N)$, another
is a directed path connecting $(I+1,1)$    and $(N,J)$ .
Please note that the former two  points should be should be the
nearest neighbors of the latter two, respectively, while the end
points of
a diagonal are in fact next-nearest neighbor.
The length of
every direced path connecting
two end points of a diagonal is $2N-1$. The length of a circular path
is
$2N$.
Q.E.D.

{\em Theorem 2.-The necessary condition of  formation of
any possible jamming configuration is $p\,\geq\,(1+p_{s}/p_{l})/N$,
where
$p_{s}$ and $p_{l}$ are respectively the smaller and larger one of
$p_{\uparrow}$ and $p_{\rightarrow}$.
This is also
sufficient for that jamming configuration can form.}

{\em Proof.}-Suppose there
is a circular path made up of only the arrows with larger density,
and there are no other arrows of this kind.
The other kind of arrows will surely be blocked by this circular
path.
Therefore $N_{l}\,=\,N$, $N_{s}\,=\,(p_{s}/p_{l})N$.
Thus  $(1+p_{s}/p_{l})/N$ is the smallest density for the jamming
configuration where the circular path is made up of one kind of
arrows.

For a circular path made up of both kinds of arrows, the density is
$2/N$, which is not smaller than the value obtained from circular
path made up of only the arrows with smaller density. They are equal
when
$p_{\uparrow}\,=\,p_{\rightarrow}$.
Q.E.D.

{}From Secs. III and IV, we obtain the upper and lower critical
densities.
When the density is smaller than $(1+p_{s}/p_{l})/N$, there can only
be moving configurations. There can only  be jamming configurations
when the density is larger than
$1/2+(N-4)/2N^{2}$  if $N$ is even, or $1/2-1/N^{2}$ if $N$ is odd.
For a density between these two critical values, both moving and
jamming configurations are possible depending on the initial
configurations.
Whenever the initial configuration is given, the final stationary
configuration is determined.

\section*{V. TYPICAL FORMATION OF JAMMING CONFIGURATION}
Although we have obtained the two critical densities, it was found
from
the simulation that a jamming transition occurs at a third critical
density below the upper critical
density.  This is because that the number of moving configurations
with density
larger than the third critical density are negligible compared with
the
jamming configurations with the same density, especially when
asymptotic
configurations are studied.
Considering the typical case for the formation
of  a jamming configuration, we determine this third critical density,
denoted as $p_{c}(N)$.
This definition of $p_{c}(N)$ is different from  and more
resonable than that in Ref. \ \cite{biham}, where it is defined to be
at
the center of the region where  moving and jamming configurations are
both non-negligible.

The jammming transition occurs soon after a circular path forms. The
cirular
path, of course, typically consists of
both up and right arrows. By ``typically'',
we mean that it has the possibility near to $1$ while the other
possibility is
very small. The circular path blocks the neighboring right arrows on
its lefthand side, or say, upside, while blocks the neighboring up
arrows on its
downside, or say, righthand side. These blocked arrows block other
arrows further.
Consequently, a global cluster with directed branching structure
emerge.
Here branching means that there are both an up and  a right arrows
connected
to a same arrow, which  belongs to  the the higher branching level.
The highest level is the circular path.

In the final jamming cluster, there are some end-arrows, which are
the
end-points of the paths, which are connected to the circular path
therefore
are the rests of the closed paths other than the circular path. For
an  end-arrow, if  it is an up (right) arrow,
there
must be no right arrow on its lefthand side and there is no up  arrow
on its
downside, while its upside (righthand side) must be occupied. So the
 density of end-arrows are
\begin{equation}
\rho_{e}(p)\,=\,p^{2}(1-p_{\rightarrow})(1-p_{\uparrow})\,
=\,p^{2}\,-\,p^{3}+p^{2}p_{\rightarrow}p_{\uparrow}\,\approx\,p^{2}.
\end{equation}
On the other hand, the average number of
end-arrows connected through paths to a arrow on
the circular path is just the
average  branching levels starting from that arrow, as a function of
$N$,
it is denoted as $b(N)$. Therefore,
\begin{equation}
\rho_{e}(p=p_{c})N^{2}\,=\,2Nb(N).
\end{equation}
It is expected that
\begin{equation}
b(N)\,\sim\,N^{1+2\alpha},
\end{equation}
from which one obtain
\begin{equation}
p_{c}(N)\,=\,CN^{\alpha},
\label{eq:alp}
\end{equation}
where $C$ is a coefficient, while $\alpha$ is the exponent.

The simulation results can be used to test the above result and
determine
$\alpha$ and $c$. With the approximate values of $p_{c}(N)$ for $N\,
=\,16,\,32,\,64,\,128,\,512$ observed from the values of p at which
the ensemble average of velocity defined there \cite{exp2} begin to be
 almost 0 on Fig. 3 in Ref. \cite{biham},
we obtain a good fit to Eq. (\ref{eq:alp}) with $\alpha\,=\,-0.14$ and
$C\,=\,0.76$.

Eq. (\ref{eq:alp}) suggests that the jamming cluster at $p_{c}$ is
fractal
with dimensionality $2+\alpha\,=\,1.86$, which is near to 91/48,
the fractal dimension of
the infinite cluster of
two-dimensional
percolation \cite{feder}. This is reasonable since the jamming
cluster forms
soon after the circular path forms, which is simailar to percolation
in this respect. But
surely the jamming cluster is fractal only when the density is near
to
$p_{c}$. Therefore we suggest that the jamming transition at
$p_{c}$ might be a second-order phase transition. From lemma 5, we
know
that the order parameter is the probability that an arbitarily chosen
arrow belongs to a closed path. This is similar to that the order
parameter of percolation is the probability that an arbitrariely
chosen
occupied site or bond belongs to an infinite cluster. Extensive
studies
concerning this issue is anticipated.

\section*{VI. DISCUSSIONS}
An intuitive analogy of the dynamics of this system is the motion of
a ball on
a structure with two valleys of different depth in gravitational
field. This structure is smooth enough for the ball to roll but there
is
friction.
When  it is put initially low enough on this structure, the ball can
only
roll to the lower valley.  The structure can be made so that for high
enough
initial position, the ball   can only roll to the higher valley. For
intermidiate height, it will roll to either of the valleys depending
on
the initial position.

The two-dimensional traffic-flow model we have studied  is  in fact a
closed system, almost so is the traffic  within a city. Therefore the
so-called ``self-organization'' is mainly the tendency to
equilibrium, with
the special pattern determined by the dynamical rules. We have seen
that
the two kinds of attractors are both stationary states. By a proper
definition of
entropy or free energy, the dynamical processes might be formalized
to processes
minimizing the free energy, the landscape of which has two minima.

For an intermidiate value of density between upper and lower critial
values,
the phase space decomposes into
 unconnected  components, since one
attractor is reached given the initial configuration.
For a certain density, the trajectory in each component is
unidirectional
and
irreversible, ending in an attractor. The attractor is either a fixed
point or two fixed points, between which the system keeps hopping.
Concerning the
evolutionary process, the dynamics is of  ergodicity
breaking\cite{exp3}.
So the ensemble average of velocity in Ref. \cite{biham} has no
meaning as
a velocity, it is nothing but the percentage of moving configurations
among all possible configurations. Ensemble average of velocity
cannot be done over all
configurations. The velocity in the asymptotic configuration is
either
$1$ or $0$, belonging to disconnected components of the phase space,
any attempt in giving an ensemble average velocity between $0$ and
$1$ is
meaningless.

We may observe some resemblance between this model and directed
sandpile model [10-12]. This traffic model can be viewed as a sanpile
model
with the addition that there are two kinds of particles (arrows) with
different toppling direction.
The critical slop is $1$ so that the arrow hops to the next only if
there is no arrow there. Similar to the original sandpile model
\cite{obukhov}, there
is nonlinear interaction of the Goldstone modes due to gradient-
dependence,
{\it i.e.} $k$-dependence in momentum representation.
The leading term of the intereaction can be expected to be
proportional to $k_{x}+k_{y}$.
But there is a great difference that
it is a closed system, while
the original sanpile model is an  open system. In the trffic-flow
model, the total number of
arrows  and even the number of up (right) arrows in each
column (row) are conserved all the time from beginning. There is no
exchanges
with enviornment. In the original sandpile model, there are
continuous
exchanges with the enviornment, the conservation is kept only after
the
critical state is arrived. This critical state is {\em metastable},
while
the traffic-flow reaches one of the {\em stable} configurations,
which are
separated by an infinite barrier.
The fractal structure exhibits only in
the jamming cluster formed  near $p_{c}$. This jamming transtion at
$p_{c}$ has been suggested to be a  second-order phase transition.
The criticality is not ``self-organized''
as  in sandpile model, the tuning parameter is the density $p$, the
order parameter is the probability that  an arbitrariely chosen arrow
belong to a closed path.
For larger
density the cluster must be compact. This can also be understood
considering the interactions of Goldstone modes. For larger $p$,
the interaction drastically reduces. Since in the final asymptotic
configuration, all the arrows are moving or stucked, this model can
also be
viewed as providing an interesting mechanism for synchronization.

\section*{ACKNOWLEDGEMENT}
This work was initiated in reading Refs. [2,4,7]  provided by the
authors.
Hoi Fung Chau and Ruibao Tao are acknowledged for stimulating
discussions. Chau is particularly thanked for critically reading the
manuscript.
This work is supported
by China Postdoctoral Science Foundation.

\newpage
\begin{picture}(200,400)
\setlength{\unitlength}{1pt}
\put(60,10){\line(1,0){150}}
\put(60,10){\line(0,1){150}}
\put(60,160){\line(1,0){150}}
\put(210,10){\line(0,1){150}}
\put(60,160){\line(0,1){150}}
\put(60,310){\line(1,-1){150}}
\put(60,130){\line(1,-1){120}}
\put(60,190){\line(1,0){120}}
\put(60,160){\line(1,-1){150}}
\put(40,304){$S'$}
\put(40,184){$O'$}
\put(40,154){$P$}
\put(40,124){$S$}
\put(40,4){$O$}
\put(220,4){$R$}
\put(220,154){$Q$}
\put(180,-5){$T$}
\put(198,184){$T'$}
\end{picture}

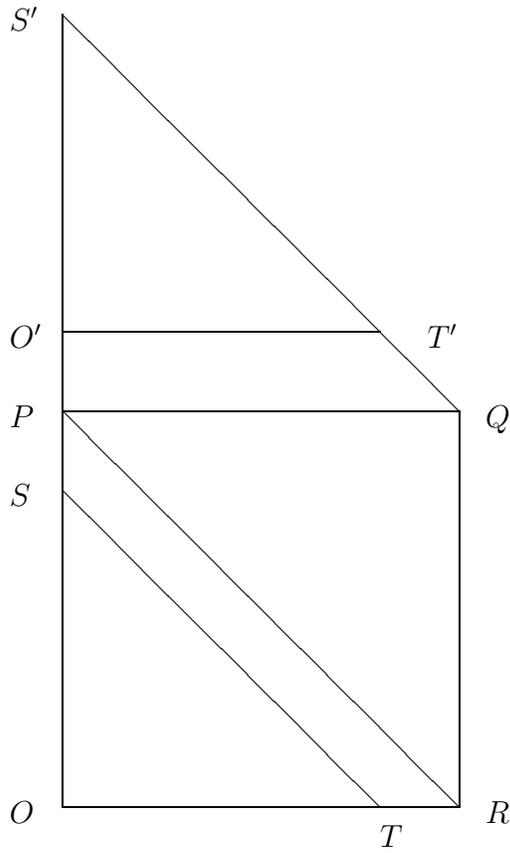
FIG. 1. Equivalent transformation of the appearance of the lattice.
The square $PORQ$ can be changed to the parallelogram $PRQS$. The
coordinates of the marked points are  $O(1,1)$, $R(N,1)$,
$Q(N,N)$, $P(1,N)$, $S(1,N-1)$, $T(N-1,1)$, $O'(1,N+1)$,
$T'(N-1,N+1)$, $S'(1,2N-1)$.
$O$ and $O'$, $T$ and $T'$, $S$ and $S'$ are
respectively the same points due to the periodic boundary condition.

\newpage

FIG. 2. Log-log plot of $p_{c}(N)$, the critical density
at which typical jamming configuration begins  to form  while the
possibility of moving configuration is negligible, with lattice size
$N$.
The circles are the results   observed from Ref. [3]. The straight
line is the least square fit yielding
$p_{c}(N)\,$$\approx\,CN^{\alpha}$,
with $C\,=\,0.76$, $\alpha\,=-0.14$.

\begin{thebibliography}{nagatani}
\bibitem{nagel} K. Nagel and M. Schreckenberg, J. Phys. (France) I
{\bf 2},
2221 (1992).
\bibitem{nagel2} K. Nagel and S. Rasmussen, in {\em Proceedings of
Alife
4}, edited by R. Brooks and P. Maes (MIT Press, Cambridge, MA, 1994).
\bibitem{biham} O. Biham, A. A. Middleton and D. Levine, Phys. Rev. A
{\bf 46}, 3290 (1992).
\bibitem{freund} J. Freund and T. P\"{o}schel, Physica {\bf A} (to
appear);
LANL e-print adap-org 9505001.
\bibitem{exp} The definition of $\tau$ here is different from that
in
Ref.\cite{biham}, where $\tau$ is the ``real'' time interval, the sum
of the
alternative odd and even time steps.
\bibitem{nagatani} T. Nagatani, J. Phys. Soc. Jpn, {\bf 62}, 2656
(1993);
Y. Ishibashi and M. Fukui, {\it ibid}, {\bf 63}, 2882 (1994).  There
are
also other defects in these papers.
\bibitem{chau} H. F. Chau, P. M. Hui and Y. F. Woo,
  preprint IASSNS-HEP-95/07; LANL e-print adap-org 9502002.
\bibitem{exp2} There are issues to be clarified in Sec. VI concerning
this
quantity.
\bibitem{exp3} There is nonrigorousness here since usually
ergodicity breaking
is for components separated by {\em finite} free energy barrier hence
can
be recovered, while it is impossible here. However, we feel that
``ergodicity
breaking'' is more suitable than ``symmetry breaking''.
\bibitem{bak} P. Bak, C. Tang and K. Wiesenfeld, Phys. Rev. Lett.
{\bf 59},
381 (1987).
\bibitem{dahr} D. Dhar and R. Ramaswamy, Phys. Rev. Lett. {\bf 63}
1659 (1989).
\bibitem{shi} Y. Shi and C. Gong, Comm. Theor. Phys. {\bf 19} 157
(1993).
\bibitem{obukhov} S. P. Obukhov, Phys. Rev. Lett. {\bf 65} 1395
(1990).
\bibitem{feder} D. Stauffer, {\em Introduction  to Percolation
Theory} (Taylor \& France, London, 1985);
J. Feder, {\em Fractals} (Plenum Press, New York and
London, 1988).
\end{thebibliography}
\end{document}